\documentclass[article,twocolumn,superscriptaddress]{revtex4-1}

\usepackage{amsfonts}
\usepackage{amssymb}
\usepackage{amsmath}
\usepackage{amsthm}

\usepackage{stackrel}
\usepackage{color}

\usepackage[pdftex]{graphicx,hyperref}

\usepackage{hyperref}

\newcommand{\bra}[1]{\langle#1|}
\newcommand{\ket}[1]{|#1\rangle}
\newcommand{\braket}[2]{\langle#1|#2\rangle}

\newcommand{\ZZ}{\mathbb{Z}}

\newcommand{\up}{\uparrow}
\newcommand{\down}{\downarrow}

\begin{document}

\title{Many-body Localization and Symmetry Protected Topological Order}

\author{Anushya Chandran}
\affiliation{Department of Physics, Princeton University, Princeton, NJ 08544, USA}
\affiliation{Perimeter Institute for Theoretical Physics, 31 Caroline Street N, Waterloo, Ontario, Canada N2L 2Y5}

\author{Vedika Khemani}
\affiliation{Department of Physics, Princeton University, Princeton, NJ 08544, USA}

\author{C. R. Laumann}
\affiliation{Perimeter Institute for Theoretical Physics, 31 Caroline Street N, Waterloo, Ontario, Canada N2L 2Y5}
\affiliation{Department of Physics, Harvard University, Cambridge, MA 02138, USA}
\affiliation{Department of Physics, University of Washington, Seattle, Washington 98195, USA}

\author{S. L. Sondhi}
\affiliation{Department of Physics, Princeton University, Princeton, NJ 08544, USA}

\date{\today}

\begin{abstract}
Recent work shows that highly excited many-body localized eigenstates can exhibit broken symmetries and topological order, including in dimensions where such order would be forbidden in equilibrium.
In this paper we extend this analysis to discrete symmetry protected order via the explicit examples of the Haldane phase of one dimensional spin chains and the topological Ising paramagnet in two dimensions.
We comment on the challenge of extending these results to cases where the protecting symmetry is continuous.
\end{abstract}

\maketitle


\section{Introduction}
\label{Sec:Introduction}
The long conjectured phenomenon of many-body localization (MBL) \cite{Anderson:1958p7531} has been put on a much firmer basis by the work of Basko, Aleiner and Altshuler (BAA) \cite{Basko:2006hh,Basko:2006ly} and a set of following investigations \cite{Oganesyan:2007uq,Pal:2010zr,Bardarson:2012qa,Iyer:2013kl,Huse:2013tx,Huse:2013ys,Bahri:2013vn, Vosk:2013fu,Rigol:2008bh, Bauer:2013ghi,Serbyn:2013abcd}.
For a range of energy densities above the ground state, the highly excited eigenstates of sufficiently disordered quantum Hamiltonians with locally bounded Hilbert spaces exhibit a set of interlinked properties:
(i) these states fail to satisfy the eigenstate thermalization hypothesis (ETH) \cite{Deutsch:1991ve,Srednicki:1994qf} so that the canonical ensemble and temperature are no longer meaningful;
(ii) the long wavelength thermal conductivity vanishes within this energy range;
(iii) neighboring eigenstates in the many-body spectrum differ significantly in their local properties; and,
(iv) the entanglement entropy of macroscopic domains is sub-extensive.
This last point should be contrasted with usual extended states wherein the entanglement entropy scales with the volume of the domain and ought agree with the canonical thermodynamic entropy.

While the above complex of properties is  generally applicable to any MBL phase %
\footnote{Here, a `phase' corresponds to a region in the energy density and parameter space, which mimics the terminology for equilibrium systems.
Readers will keep in mind that the system is by definition {\it not} in equilibrium in an MBL phase.},
recently Huse et al \cite{Huse:2013tx} observed that the eigenstates could be more finely classified with reference to various measures of order.
MBL eigenstates can spontaneously break or preserve global symmetries and exhibit or fail to exhibit topological order.
These phenomena could violate the naive expectation from Peierls-Mermin-Wagner type arguments.
Essentially, the localization of defects allows order to persist at energy densities where equilibrium arguments predict destruction of order.

In this article we extend the analysis of Ref.~\onlinecite{Huse:2013tx} to a case intermediate between symmetry-breaking  and topological order.
This is the case of symmetry protected topological order (SPT) \cite{Gu:2009dq,Chen:2011cr,Fidkowski:2011tg,Schuch:2011kl}, wherein a symmetry is needed for the phase to exist but the order itself is topological in nature and cannot be characterized by a local order parameter.
Clean zero temperature SPT phases have a bulk gap to well-defined excitations whose quantum numbers are not fractional.
Furthermore, SPT ground states cannot be continuously connected to trivial product states without either breaking the protecting symmetry or closing the energy gap; however, such a continuous path must exist if the protecting symmetry is explicitly broken.
The canonical example of an SPT phase is the Haldane phase in $d=1$ \cite{Haldane1983464,PhysRevLett.50.1153} and the most celebrated one is by now surely the $Z_2$ topological insulator in $d=3$ (reviewed in Ref.~\onlinecite{Hasan:2010hc}).

With this background, we can now state our central question: Can highly excited eigenstates exhibit SPT order in the presence of MBL? We take such order to
generalize the cluster of properties listed above. Specifically, we wish
to examine Hamiltonians invariant under a protecting symmetry with highly excited eigenstates that lie in a mobility gap.
We will require an eigenstate phase transition (at which the properties
of the eigenstates change in some singular fashion) between the SPT
region and the trivial region, which is well captured by product states as
long as the protecting symmetry is intact.
Further, there should be a path along which such a phase transition is absent when the symmetry is explicitly
broken.

In the following, we address this question via two examples.
The first is the Haldane phase protected by a discrete symmetry.
We present strong evidence that the SPT order extends in an MBL version to highly excited eigenstates even though equilibrium considerations preclude such order.
We do so by introducing an appropriate generalization of the AKLT model of Affleck, Kennedy, Lieb and Tasaki  \cite{PhysRevLett.59.799,affleck1988vbg} that allows the arguments of BAA to be brought to bear on highly excited states.
We discuss various diagnostics of the Haldane phase that extend to this regime.
We also note that the Haldane phase with continuous $SU(2)$ symmetry does not obviously extend to an MBL version and explain the obstacles involved in settling this question.
Our second example is the topological Ising paramagnet in two dimensions \cite{Levin:2012jq,Chen:2011ij}.
Here again we adapt the BAA arguments to establish MBL and
discuss the diagnostics needed to establish SPT order.
We conclude with some comments on generalizations and open questions.

As we were finishing this article, we became aware of the preprint, Ref.~\onlinecite{Bahri:2013vn}. In this preprint, the authors study MBL in a one dimensional spin-1/2 model related to our first example, the Haldane phase, from the perspective of edge modes, the entanglement spectrum and string order. We discuss the precise connection between our work and theirs at the end of Sec.~\ref{Sec:Haldane}.

\section{Haldane phase}
\label{Sec:Haldane}
\subsection{Review of low energy physics}
We begin with the Haldane phase of the spin-1 antiferromagnetic chain.
Although usually understood in the context of continuous rotational symmetry
\footnote{It is interesting to note that Haldane discovered the phase that bears his name studying a dihedral-symmetric perturbation of the SU(2) invariant spin chains.}, the Haldane phase is an SPT which may be protected by any one of the following discrete symmetries: inversion, time reversal or the dihedral group $D$ of $\pi$-rotations around the $x,y,z$-spin axes \cite{Berg:2008fk, Pollmann:2010ih,Pollmann:2012nx}.
At zero temperature, the clean phase is a gapped quantum spin liquid which breaks none of these symmetries.
It has several defining characteristics.
First, the bulk exhibits simultaneous long-range ``string'' order \cite{denNijs:1989dp, PhysRevB.45.304} in the operators $(\alpha = x,y,z)$
\begin{align}
	\label{eq:stringorderparameter}
	\sigma^{\alpha}_{ij} = - S_i^\alpha \left(\prod_{k=i+1}^{j-1} R_k^\alpha \right) S_j^\alpha
\end{align}
where $R_j^\alpha = e^{i\pi S^\alpha_j}$ represents a rotation by $\pi$ around the $\alpha$ spin axis of site $j$ and $S^\alpha_i$ are the usual spin-1 operators.
Second, the boundary exhibits protected spin-$1/2$ edge modes as a consequence of which the ground state is four-fold degenerate on open chains.
Third, the presence of the protected spin-$1/2$ edge modes implies a two-fold degeneracy in the entanglement spectrum for virtual (Schmidt) cuts in the bulk of the chain.
Further, the underlying spin-1 degrees of freedom do not fractionalize in the bulk, in consonance with the definition of an SPT. The low energy excitations are gapped spin-1 bosons called `triplons', discussed later in the text.
In contrast, in the trivial phase with the same discrete symmetries, the ground state can always be smoothly connected to a product state through a symmetric path\cite{Pollmann:2010ih}. The trivial phase has no string order, boundary modes or degenerate entanglement spectra; hence these properties signal the SPT order of the Haldane phase.

\subsection{Ergodicity and localization in highly excited states}
In the following, we will review how these signatures of the Haldane phase disappear at $T>0$ in clean systems as a consequence of the delocalization of the triplons in highly excited states.
On the other hand, in the presence of sufficient disorder, we will argue that individual triplons Anderson localize.
At sufficiently small, but non-zero, energy density, the dilute gas of localized triplons interacts only weakly so that the perturbative arguments of BAA apply and the system is many-body localized.
Finally, we will discuss how various defining characteristics of the Haldane phase persist to finite energy density in a suitably modified form in this MBL phase.

To be concrete, we introduce a frustration-free model for the Haldane phase.
As the SPT order requires only the dihedral group $D = \{\mathbf{1},R^x,R^y,R^z\} \equiv \ZZ_2 \times \ZZ_2$ to protect it, our model has precisely this symmetry, but is otherwise very closely related to the celebrated $O(3)$-symmetric AKLT model \cite{PhysRevLett.59.799,affleck1988vbg}.
The Hamiltonian, which we refer to as the BKLT Hamiltonian,  is
\begin{widetext}
\begin{align}
	\label{eq:bklt_ham}
	H_{BKLT} = \sum_{i,\alpha} P^{(2)}_{i,i+1} \left( J_i + c_i^\alpha (S_i^\alpha + S_{i+1}^\alpha)^2 + d_i^\alpha (S_i^\alpha + S_{i+1}^\alpha)^4 \right) P^{(2)}_{i,i+1}
\end{align}
\end{widetext}
where $P^{(2)}_{i,j}$ projects onto the spin-2 representation of the spins $i$ and $j$, and $J_i , c_i^\alpha, d_i^\alpha>0$ are coupling constants \footnote{If \unexpanded{$c_i^\alpha, d_i^\alpha>0$}, then each term is strictly positive. Note that in the spin-$2$ representation, the coupling constants are not all independent as $(S_i^\alpha + S_{i+1}^\alpha)^2=6$. Taking \unexpanded{$c_i^\alpha, d_i^\alpha$} to zero reduces BKLT to the traditional AKLT model. }.
The ground state space of $H_{BKLT}$ is identical to that of the AKLT model:
there are four ground states on open chains, each of which possesses an explicit, compact matrix product state (MPS) representation simultaneously annihilated by all $P^{(2)}_{i,i+1}$ and therefore by $H_{BKLT}$.
The excitation gap is of order $J_i$ and the eigenstates may be labeled by the one-dimensional representations of $\ZZ_2\times\ZZ_2$.
Even though the ground states are exactly known, $H_{BKLT}$ is not fully integrable. Its excited states should therefore be generic with respect to thermalization and many-body localization.

The A/BKLT ground states can be constructed by splitting each spin $1$ site into two virtual spin $1/2$ degrees of freedom. Pictorially,
\begin{align}
	\ket{A;v_L, v_R} = \vcenter{\hbox{\includegraphics[scale=0.8]{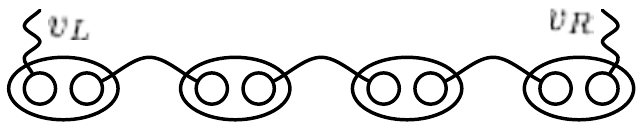}}}
\end{align}
where each small circle represents a virtual spin $1/2$, the solid lines denote singlet pairings and the ovals the symmetrization to reproduce a spin $1$ physical degree of freedom.
Here, $v_L$ and $v_R$ are the state vectors for the boundary spins that label the four-dimensional ground state space on the open chain.
This picture immediately reveals the physical origin of the spin $1/2$ boundary modes -- they correspond to the unpaired virtual degrees of freedom left on either end of the open chain.
The picture also suggests the origin of the $2$-fold degeneracy in the entanglement spectrum as the cutting of the virtual Bell pair shared by a link.

The virtual spin structure of the A/BKLT state suggests a natural candidate for the low energy bulk excitations,
\begin{align}
	\label{eq:single_triplon}
	\ket{j, \alpha} =
	\vcenter{\hbox{\includegraphics[scale=0.8]{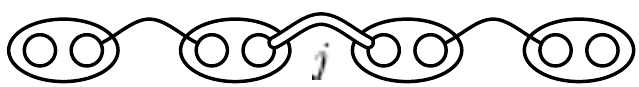}}}
\end{align}
where the double line at bond $j$ indicates a virtual pair in triplet state $\alpha$. Note that we have suppressed the explicit boundary spin states $v_L, v_R$.
The single `triplon' states $\ket{j,\alpha}$ are non-orthogonal but linearly independent.
They span the manifold studied in the single-mode approximation (SMA) provided by $S^\alpha_j$ operators acting on $\ket{A}$ \cite{PhysRevLett.60.531}.
\footnote{On a periodic chain of length $L$, there are $L$ linearly independent bond triplons as we have defined them. The spin operators $S^{\alpha}_j$ create superpositions of the form {$S^{\alpha}_j \ket{A} \propto \ket{j,\alpha} - \ket{j-1,\alpha}$} and thus there are only $L-1$ linearly independent states in the traditional SMA calculation.}
These states are believed to be good variational approximations to the local excitations of $H_{AKLT}$, in part because the SMA calculations produce a single triplon band quantitatively in good agreement with numerical studies \cite{PhysRevLett.60.531}.
We note in passing that the bond triplon states provide a superior framework for the study of excitations in higher dimensional valence-bond solid states as well, where the SMA is inadequate.

In the $O(3)$ symmetric AKLT case, the three triplon states $\ket{j,\alpha}$ are strictly degenerate. Breaking the $O(3)$ symmetry down to the dihedral subgroup, as in BKLT, lifts the degeneracy and selects the dihedral-symmetric states $\ket{x}, \ket{y}, \ket{z}$ as an appropriate basis.
In terms of virtual spins,
\begin{align*}
	\ket{x} &= (\ket{\up\up} - \ket{\down\down})/\sqrt{2} \\
	\ket{y} &= (\ket{\up\up} + \ket{\down\down})/\sqrt{2} \\
	\ket{z} &= (\ket{\up\down} + \ket{\down\up})/\sqrt{2}
\end{align*}
where $\ket{x}$ has eigenvalues $+1, -1, -1$ under $R^x, R^y, R^z$, $\ket{y}$ has $-1, +1, -1$ and $\ket{z}$ has $-1, -1, +1$.
The reader should recognize that dihedral symmetry has picked out the maximally entangled Bell states!

Consider now the three diagnostics of the Haldane phase in the presence of a maximally localized triplon.
(i) As the virtual spins in $\ket{j,\alpha}$ form a Bell state across every bond, the entanglement spectrum exhibits two-fold degeneracy across any real space cut.
It is straightforward to confirm this using the explicit MPS representation of $\ket{j,\alpha}$ following from Eq.~\eqref{eq:single_triplon}.
(ii) The triplon excitation produces a topological defect in the string order parameter $\sigma^\beta_{ik}$.
Explicitly,
\begin{align}
\label{Eq:OneTripStringOrder}
	\bra{j, \alpha} \sigma^\beta_{ik}\ket{j,\alpha} = \left\{ \begin{array}{ll}
	-(-1)^{\delta_{\alpha\beta}}\frac{4}{9} & i \le j < k \\
	\frac{4}{9}	& \textrm{else}
	\end{array}\right.
\end{align}
That is, if the string operator crosses the triplon, it picks up a minus sign unless the flavors of the string and the triplon agree.
(iii) On open chains, there remain four degenerate, linearly-independent variational states corresponding to the choice of boundary conditions ($v_L, v_R$) for the localized triplon state $\ket{j,\alpha}$ %
\footnote{The issue of linear independence for bond triplon states is somewhat delicate. On an open chain of length $L$, there are naively $12(L-1)$ triplon states corresponding to the 4 boundary states, 3 triplon flavors and $L-1$ positions. These span only a $12(L-1) - 4$ dimensional space. On a closed chain, there are $3L$ linearly independent states.}.

The demise of the Haldane phase at finite energy density in the clean system is now apparent.
Diagonalizing $H_{BKLT}$ in the variational single triplon manifold gives rise to three delocalized bands of triplons corresponding to each of the flavors $\alpha$.
This follows from solving the generalized eigenvalue problem where $H_{BKLT}$ is purely diagonal in the localized triplon basis while the overlap matrix $\braket{j,\alpha}{k,\beta} \sim \delta_{\alpha\beta} (1/3)^{|j-k|}$ produces the off-diagonal dispersion.
At low energy densities, we expect a dilute gas of these delocalized triplons in the eigenstates of $H_{BKLT}$.
This fluctuating gas (i) produces an extensive entanglement entropy for macroscopic domains which precludes an MPS representation for the highly excited eigenstates and washes out the two-fold entanglement degeneracy.
(ii) As the triplons act as defects in the string order Eq.~\eqref{eq:stringorderparameter}, their spatial fluctuations suppress this order on the length scale of the inverse density.
Finally, (iii) the spin-1/2 boundary modes decohere due to interaction with the delocalized bulk triplons on a time scale set by the density of triplons.
This is all consistent with the expectation that there is no order, topological or otherwise, at finite temperature in one dimension.

\begin{figure}[tbp]
	\centering
		\includegraphics{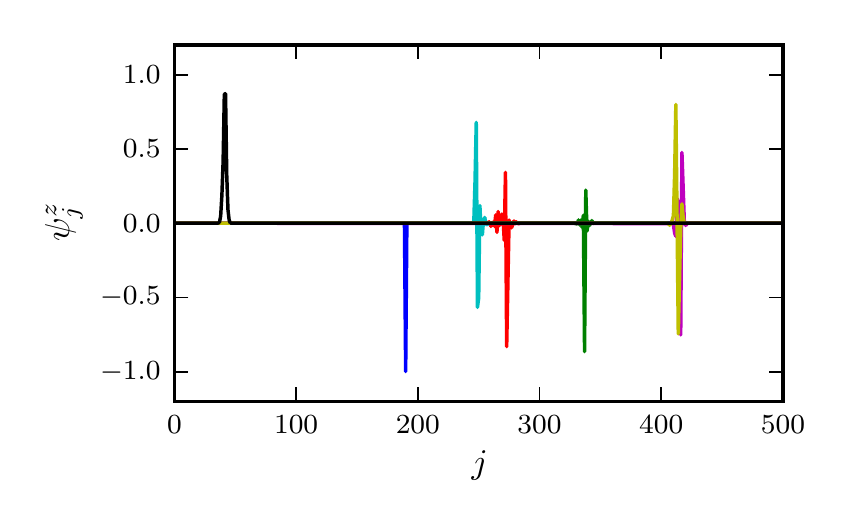}
	\caption{Seven typical eigenmodes of the Anderson problem in the single $\alpha=z$ triplon manifold in a $500$-site chain with periodic boundary conditions. The coupling constants $J_i$ are drawn uniformly from the interval $\left(0,1\right)$.  }
	\label{fig:anderson_WF}
\end{figure}

The presence of sufficient disorder leads to an entirely different picture of the highly excited eigenstates -- that they may many-body localize and thus retain their SPT character.
Consider the introduction of disorder in the couplings of $H_{BKLT}$.
So long as $J_i> 0$, the ground state is completely unperturbed by this variation, which is an extreme manifestation of the insensitivity of gapped phases to weak spatial disorder.
The excitation spectrum, on the other hand, changes dramatically.
Even for weak variations $\delta K \ll K$ for $K=J, c, d$, we expect the single triplon eigenstates to Anderson localize. This follows from analyzing the generalized eigenvalue problem described in the paragraph above with spatially varying diagonal matrix elements.
Fig.~\ref{fig:anderson_WF} shows the typical localized triplon wavefunctions found by this analysis.

We now make the case for MBL following BAA.
Consider the excited states with a low density of localized triplons.
The interaction $U$ between two triplons separated by a distance $l$ scales as $J e^{-l/\xi}$, where $\xi$ is the longer of the triplon overlap decay length ($1/\log(3)$) and the localization length.
When the typical spacing $l$ between excitations is sufficiently large so that the typical energy splitting between nearby states (of order 1) is much larger than the interactions, $U \sim \pm J e^{-l / \xi}$, the perturbative BAA arguments protect triplon localization.
That is, the system remains many-body localized up to a finite energy density $\epsilon$ such that the typical separation $J/\epsilon$ is small on the scale $\xi$.

\begin{figure}[tbp]
	\centering
		\includegraphics{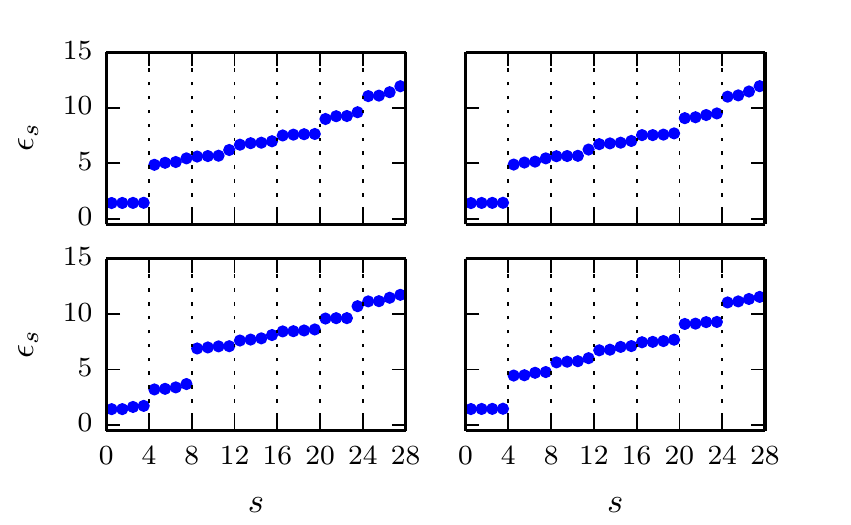}
	\caption{The entanglement spectra of four consecutive excited states, starting with the $60$th state above the ground state, for a 12-site open BKLT chain with disorder in all coupling constants. Each state is decomposed into two equal halves.}
	\label{fig:entspec_4}
\end{figure}

The naive application of the same argument fails as one approaches the $O(3)$ symmetric AKLT point by taking $c_i, d_i$ to zero.
In this limit, the local fields splitting the triplet degeneracy vanish so that there is no regime where the typical interaction strength $U$ is smaller than the typical local level spacing.
Rather, the localized triplons carry spin-1 and the system of a dilute random array of non-interacting triplons is highly degenerate.
From this point of view, the interactions (still of order $U \sim J e^{-l / \xi}$) split this large degeneracy according to a disordered system of both ferro- and antiferromagnetic exchanges.
Whether such an effective $O(3)$-symmetric random spin-1 chain can exhibit a MBL phase is an intriguing open question.
The application of the real space renormalization group to such a system suggests that the system ought to grow large effective moments \cite{Westerberg:1995pi, Hyman:1997cu} which, if they behave classically as one expects of large spins, would lead to thermal conduction and equilibration \cite{Oganesyan:2009ff,Basko:2011lh}.

Finally, we consider how the three signatures of the Haldane phase persist to finite energy density in the MBL regime.
First, take the `caricatures' of the excited states at low energy density given by the MPS with a low density of double lines at prescribed bonds as in Eq.~\eqref{eq:single_triplon}.
We have already noted that (i) the entanglement spectrum is doubly degenerate,
(ii) the string order is `glassy' (Eq.~\eqref{Eq:OneTripStringOrder}), and
, (iii) the expectation value of $H_{A/BKLT}$ is independent of the virtual spins $v_L$, $v_R$ on the boundary.
Thus, if the `caricature' states were the true excited eigenstates in the presence of disorder, all of the characteristics of the Haldane phase would persist to low energy density.

Of course, the simple caricatures neglect the `fuzziness' in the position of the triplons in Anderson localized single particle wavefunctions such as in  Fig.~\ref{fig:anderson_WF}.
To construct multi-triplon `filled' Anderson localized states, we define the bond triplon creation operators:
\begin{align}
	t_j^\alpha = \prod_{i \le j} R^\alpha_i
\end{align}
These commuting, self-adjoint, unitary operators place triplons of type $\alpha$ at bond $j$ when acting upon the A/BKLT ground state space.
The single triplon localized states are then created by
\begin{align}
\label{Eq:AndersonTrip}
	t_\psi^\alpha = \sum_j \psi^\alpha_j t_j^\alpha
\end{align}
acting on the A/BKLT vacuum, where $\psi^\alpha_j$ are the eigenmodes in the single triplon problem.
We caution that the mode functions $\psi^\alpha_j$ are not orthonormal as they are coefficients with respect to a non-orthogonal basis, and neither do the $t_j^\alpha$ satisfy a canonical algebra.
Nonetheless, for sufficiently dilute collections of triplons, we expect the Fock states $\ket{\Psi} = t_{\psi_1}^{\alpha_1}t_{\psi_2}^{\alpha_2}\cdots t_{\psi_N}^{\alpha_N}\ket{A}$ to be good approximate representations of the MBL eigenstates.
Just as localized Fock states of normal bosons and fermions have entanglement entropy satisfying an area law,  $\ket{\Psi}$ has an area law for localized $\psi^\alpha_j$.
Thus, such states can be recast to exponential accuracy as finite dimensional MPS states which in turn fall into the two-fold SPT classification of dihedral symmetric states \cite{Pollmann:2010ih,Turner:2011fj}. 
We recapitulate this argument in more detail in Appendix~\ref{Sec:AppES} for non-translation invariant states.
In the same appendix, we argue that the fuzzy Fock states above are in the same non-trivial class as the A/BKLT ground state, that is, they exhibit two-fold degenerate entanglement spectra in the bulk for a single spatial cut.
Numerical exact diagonalization results are consistent with this prediction.
In Fig.~\ref{fig:entspec_4}, we plot the entanglement spectra of a few excited states of the 12-site open BKLT chain with disorder.
Dihedral symmetry forces the physical spin halves at the two boundaries to be maximally entangled; thus the spectrum should be 4-fold degenerate if the excited state has SPT order.
There is evidence of this degeneracy in Fig.~\ref{fig:entspec_4}.
In conclusion, states such as $\ket{\Psi}$ exhibit (i) two-fold degenerate entanglement spectra in the bulk, and (ii) long-range string glass order with softened frozen in domain walls and (iii) spin-1/2 boundary modes associated with the projective representation of the corresponding finite dimensional MPS.

In the presence of dihedral symmetry, the string glass order diagnoses the non-analyticity of the eigenstates at the transitions between the SPT MBL phase and the trivial MBL phase (or the ergodic phase).
On the other hand, without dihedral symmetry, such an order parameter distinction disappears.
For example, turning on a local N\'{e}el field induces a N\'{e}el magnetization, and as shown in Ref.~\onlinecite{PhysRevB.45.304}, string order.
Thus, the non-analyticity associated with the loss of the long range string glass order will be lost and the eigenstates in both MBL phases can be smoothly connected.

We end with a few comments.
First, the recent numerical study in Ref.~\onlinecite{Bahri:2013vn} probed the boundary modes of excited MBL states in a related one-dimensional model using spin-echo.
Such numerical experiments are unavailable in the disordered BKLT model due to the large intrinsic correlation lengths as compared to accessible system sizes.
Second, a consequence of the existence of boundary modes is a `pairing' regime in the many-body energy spectrum of open chains.
In this regime, the four boundary states can be identified by their small splitting relative to the exponentially small many-body spacing \cite{Huse:2013tx}.
However, there is evidence from perturbative and numerical calculations in the non-integrable Majorana chain that this pairing may persist to the clean limit \cite{Laumann:fu}.
The relationship between pairing and coherent boundary modes is thus not settled and requires further study.
Finally, the entire discussion in this section is not special to the A/BKLT point. The MBL phase at low energy densities continues away from these points.

\section{Topological Ising paramagnet in d=2}
\subsection{Review of low energy physics}
\begin{figure}
\includegraphics[width=\columnwidth]{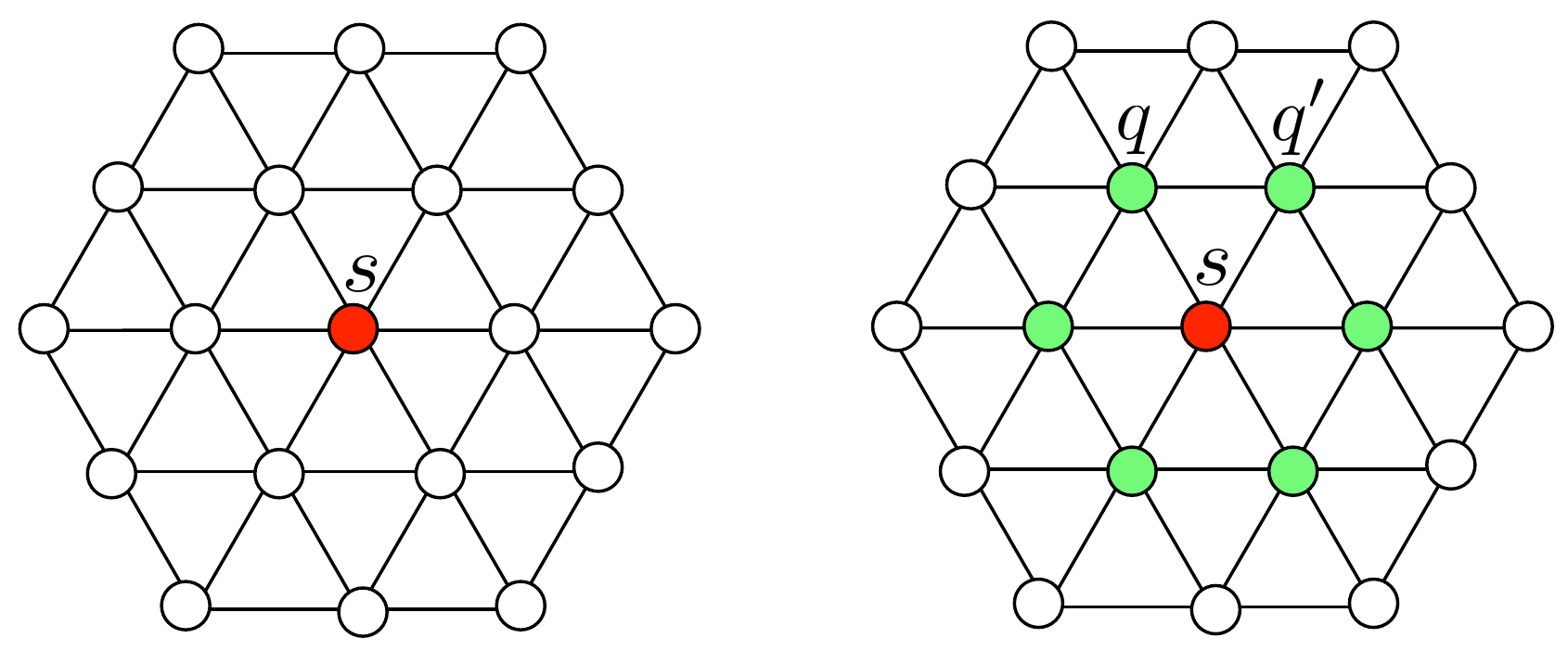}
\caption{ (Left) Trivial paramagnet $H_0$ defined in \eqref{Eq: H_0} is a sum of $\sigma_s^x$ terms on sites $s$ of the triangular lattice. (Right) Topological paramagnet $H_{LG}$ defined in \eqref{eq:LGHam} is a sum of seven-spin terms involving a product of  $\sigma_s^x$ and a phase factor $\prod_{\langle s q q'\rangle} i^ {\frac{1- \sigma_q^z \sigma_{q'}^z}{2}}$ from the six spins surrounding $s$.  }
\label{fig:LGHam}
\end{figure}

We now turn to discrete SPT phases in higher dimension.
In particular, we consider two dimensional spin systems with $\mathbb{Z}_2$ symmetry, where there is a two-fold classification of SPTs:
the trivial and the topological Ising paramagnets \cite{Levin:2012jq,Chen:2011ij,Chen:2013bs}.
We work near an exactly solvable model in the topological SPT phase, first constructed by Levin and Gu \cite{Levin:2012jq}:
\begin{align}
H_{LG}  = - \sum_{s} \Lambda_s B_s, \qquad B_s =  -\sigma_s^x\prod_{\langle s q q' \rangle} i^{\frac{1-\sigma_q^z\sigma_q^z}{2}}.
\label{eq:LGHam}
\end{align}
Here, $\Lambda_s$ are coupling constants, $\sigma_s$ are Pauli spin-1/2 operators living on the sites $s$ of a triangular lattice, and the product in $B_s$ runs over the six triangles $\langle s q q' \rangle$ intersecting the site $s$ (see Fig.~\ref{fig:LGHam}).

The Hamiltonian is invariant under the protecting Ising symmetry $S = \prod_s \sigma_s^x$.
The $B_s$ operators on different sites commute with each other, and the gapped paramagnetic ground state is the simultaneous $B_s = +1$ eigenstate $ \forall s$.
On closed manifolds, this ground state is unique and can be written explicitly in the $\sigma^z$ basis as a superposition of all spin configurations each with amplitude $(-1)^{N_{dw}}$, where $N_{dw}$ is the number of domain walls in the configuration.
These non-trivial phase factors reflect the topological nature of the ground state.

The topological paramagnet (TPM) is to be contrasted with the better-known trivial paramagnet (TrPM) with exactly solvable model Hamiltonian
\begin{align}
H_0 = -\sum_s \Gamma_s \sigma_s^x,
\label{Eq: H_0}
\end{align}
where $\Gamma_s$ are coupling constants.
The ground state of the trivial paramagnet is clearly a simple product state which, in the $\sigma^z$ basis, corresponds to a uniform superposition of all spin configurations with amplitude $1$.

The excitations in both models correspond to `spin flips' which are sites $s$ with either $B_s = -1$ or $\sigma^x_s = -1$, respectively.
At the exactly solvable points, such spin flips are static and thus the highly excited eigenstates are already many-body localized analogous to the `caricature' states of the previous section.
Absent disorder, this form of MBL is non-generic: any non-commuting perturbations to the model Hamiltonians induce dispersion of the spin flips, which in turn destroys many-body localization. 
For specificity, we add a ferromagnetic coupling term to make the spin excitations dynamical and consider Hamiltonians of the form:
\begin{align}
\tilde{H}_{0/LG} = H_{0/LG} - J \sum_{\langle s s' \rangle} \sigma_s^z \sigma_{s'}^z.
\label{eq:Htilde}
\end{align}
For $J$ large enough, the ferromagnetic term drives a transition out of either paramagnet into a symmetry broken ferromagnetic phase.

\subsection{Ergodicity and localization in highly excited states}
Now include randomness in the couplings $\Lambda_s$ and $\Gamma_s$. For simplicity, keep $\Lambda_s, \Gamma_s > 0$ to preserve the exact ground state. In this regime, the individual spin flip manifold remains Anderson localized even with small `hopping' $J$.
BAA arguments suggest that dilute gases of these weakly interacting point particles remain many-body localized. 
It is intuitively clear that both paramagnets continue into MBL versions at finite energy density as the defects that would destroy the SPT order are localized.
In the following, we will consider the extension of various SPT diagnostics to finite energy density MBL states to substantiate this intuition. We first distinguish the MBL topological and trivial paramagnets from the extended thermal paramagnetic phase, and then turn to diagnostics that differentiate the two MBL paramagnets. 

The MBL paramagnets can be easily distinguished from their thermal counterparts at nonzero energy densities using the behavior of certain Wilson loops \footnote{The two paramagnets are smoothly connected when thermalized, so no distinction is necessary.}. 
Recall that in 2+1 dimensions magnetic systems with site variables are dual to gauge theories with bond variables \cite{RevModPhys.51.659}. The spin models $H_0/ H_{LG}$ are respectively dual to the perturbed toric-code (t.c.)/ doubled-semion (d.s) $\mathbb{Z}_2$ gauge theories, with the t.c/d.s theories restricted to a static matter sector. These dual gauge theories live on the honeycomb lattice; their topologically ordered deconfined phases map to the paramagnetic phases of the spin models, while their confined phase maps to the ferromagnetic phase. The doubled semion model is discussed in \cite{levin:045110}.

As the dual models are pure gauge, their respective deconfined phases may be diagnosed by the celebrated perimeter-law of equal time Wilson loops. Each of the two deconfined phases, has a (different) canonical Wilson loop which minimally probes the confinement of charges without further exciting the gauge sector \cite{levin:045110}.
The Wilson loops of the dual gauge theories correspond to the following operators in the original spin variables $\sigma_s$, :
\begin{align}
W_0[C] = \left\langle\prod_{s\in A[C]} \sigma_s^x  \right\rangle \label{eq:Wilson_0}\\
W_{LG} [C] = \left\langle \prod_{s \in A[C]} B_s \right\rangle \label{eq:Wilson_LG}
\end{align}
where the product is over all sites $s$ lying within $A[C]$, the area enclosed by the curve $C$. These Wilson loops exhibit the ``zero-law" $W_{0/LG} [C] = 1$ exactly at the pure trivial/topological paramagnetic points. The zero-law continues to a perimeter law $W[C] \propto e^{- c |C|}$ on perturbing away from the exactly solvable points. On the other hand, the Wilson loops exhibit an area law $W[C] \propto e^{-c'|A[C]|}$ in the ferromagnetic phase. 
 
For clean, ergodic systems, both Wilson loops exhibit an area law at any finite temperature $T >0$. 
This reflects the presence of a finite density of delocalized vortex excitations in the dual gauge theories. 

The problem with disorder was recently discussed for the standard $\mathbb{Z}_2$ gauge theory by Huse et. al. \cite{Huse:2013ys}. 
In the presence of sufficient randomness in the couplings of the dual gauge theory, there exists a MBL topologically ordered phase for the $\mathbb{Z}_2$ gauge theory at finite energy density.
The excited MBL eigenstates have a finite density of localized vortices, whence the Wilson loop  $W$ exhibits a ``spin-glass" version of the perimeter law --- the magnitude of $W$ decays as the perimeter of C, but with a sign that depends on the number of localized vortices enclosed by C \cite{Huse:2013ys}. 
An analogous story holds for the doubled-semion gauge theory as well. 
By duality, the MBL highly excited eigenstates of the trivial and topological
paramagnets exhibit a spin-glass perimeter law for $W_0[C]$ and $W_{LG}[C]$ respectively. 
By contrast, excited eigenstates for the thermal paramagnet exhibit area laws for these quantities just as in the clean limit. Thus, a sharp distinction exists between the MBL and thermal phases for the two paramagnets, diagnosed by the behavior of the Wilson loop operator. 

We now turn to the question of diagnosing the two MBL paramagnets as distinct phases. 
One's first instinct might be to use the Wilson loops and, for the ideal Hamiltonians, they work: 
	$W_0[C] = 1$ for the TrPM and vanishes for the TPM, while $W_{LG}[C] = 1$ for the TPM and vanishes for the TrPM. 
Unfortunately this does not hold more generally; both Wilson loops exhibit a perimeter law in both paramagnetic phases. 
Possibly the ``correct'' one is always dominant, but this is a topic for future
work. 

Instead, let us consider other possible diagnostics to separate the MBL TrPM and TPM phases. (i) At $T=0$ in the ground state, the edge of the TPM must either by gapless or break the $\mathbb{Z}_2$ symmetry. (ii) If we gauge the models, the gauged TPM exhibits vortices with semionic statistics, which, (iii) in the presence of time reversal symmetry, bind Kramers doublets \cite{Zaletel:2013ud}\footnote{We thank M. Zaletel for bringing this to our attention.}. We expect each of these properties to extend to the MBL phase, as we explore below.

The gaplessness of the symmetric edge is not a sharp diagnostic of the TPM, even at $T=0$, as already alluded to  by Levin and Gu in the clean case. 
The edges can always spontaneously gap by breaking Ising symmetry for arbitrarily weak perturbations; of course, gapped symmetry-broken edges can also be present in the TrPM.
With disorder at finite energy the situation is even worse --- the many-body spectrum is always gapless although local operators may exhibit a `mobility' gap in localized states. Thus, we might expect `mobility gaplessness' in the absence of symmetry breaking, but this is a delicate diagnostic at best.

At $T=0$, Levin and Gu proposed a sharp distinction between the two paramagnets based on a different diagnostic. 
They coupled both paramagnets to a static gauge and then considered the statistics of braiding $\pi$ flux vortex insertions. 
For the TrPM the statistics are bosonic while for the TPM they are semionic, as the gauged models are dual to the toric code and doubled semion theories, respectively.
In a putative MBL state, a slow physical process of inserting fluxes, braiding and annihilating them should accumulate the same semionic statistical phase (on top of `spin glass'-like Aharonov-Bohm contribution from each of the encircled localized charges). 
The definition of `slow' is subtle as the many-body spectrum is gapless, but again we expect a local $O(1)$ mobility gap. 
The exact mathematical operators which characterize this process in the exactly solvable models do not have simple extensions to the general MBL state.

If the gauged paramagnet additionally has time reversal symmetry, then each vortex of the TPM binds a Kramers doublet (the semion and the anti-semion states). This can be seen in the exactly solvable model by defining a local charge operator on an area $A$, $Q[A] = \prod_{p\in A} B_p$, gauging it and noting that the gauged $Q$ is time-reversal odd (even) if $A$ encloses an odd (even) number of vortices. This implies an exact degeneracy for the entire spectrum. 
On the other hand, the TrPM vortices are bosonic and do not bind Kramers doublets (the gauged charge operators are always time-reversal even in the exact model). 
The degeneracy lifts exponentially in the separation between the vortices on perturbing away from the exactly solvable point and we expect this exponential degeneracy to persist into the MBL phase. The careful reader might note that the typical many-body level spacing for highly excited states is exponentially small in the system volume, and thus smaller than the separation between paired states. This is reminiscent of the paired MBL regime discussed by Huse et. al \cite{Huse:2013ys}, and the `paired' states share  all their local properties unlike typical MBL eigenstates close in energy. 
A different but related diagnostic comes from measuring coherent `anyon oscillations' between the semion and anti-semion states in the localized background with a timescale set by their separation.

We leave the detailed mathematical understanding of these last questions as open problems for future work.

Finally, we comment briefly on the requirement that there be a continuous path connecting the MBL phases of the TPM and the TrPM if Ising symmetry is broken along the path. Levin and Gu explicitly construct a local Ising symmetry-breaking unitary operator $U(\theta)$ which transforms $H_0$ into $H_{LG}$ (with $\Lambda_s = \Gamma_s$) along a path in Hamiltonian space parameterized by the continuous variable $\theta$; the same unitary can also be used for random couplings $\Lambda_s$. 
The many-body energy spectrum, and hence the level-statistics of $H(\theta)$ are identical everywhere along the path which strongly indicates the absence of a MBL to ergodic phase transition in accordance with work done by Huse et. al. \cite{Oganesyan:2007uq}. More strongly, each localized excited eigenstate of $H_0$ continues to a localized eigenstate of $H(\theta)$ under the action of the local unitary, and there is a continuous mapping between MBL eigenstates everywhere along the path.
This is to be contrasted with the eigenstate phase transition that we expect between the TPM and TrPM highly excited eigenstates when Ising symmetry is preserved.

\section{Concluding remarks}

Traditionally, the destruction of order and the proliferation of defects are closely
intertwined in statistical mechanics. This has led previously to the idea that the
localization of defects can improve order, e.g. in the case of superconductors in
a magnetic field \cite{Huse:1992dz} and the quantum Hall effect away from the center of the plateau \cite{Jain:2007fk}.
The work of Huse et al has generated the interesting possibility that
this mechanism can operate also in many body localized quantum systems where
statistical mechanics does not apply even for highly excited eigenstates. In this setting
the sought after order has to be identified for individual many-body eigenstates
and has a ``spin glass'' form or at least a spin glass component which is eigenstate
specific.

In this paper we have considered whether SPT order can exist in highly excited eigenstates
in the MBL setting by examining two specific models. In both cases it is not hard
to see that thermal states differ qualitatively from the ground states exhibiting SPT
order while MBL states qualitatively resemble the ground states thanks to the localization of defects. This is strong evidence for existence of an eigenstate phase transition
that must separate the trivial and SPT regions at nonzero energy density.
For the case of the Haldane phase in $d=1$ we are able to go further and argue that highly excited MBL eigenstates in the SPT region can be directly distinguished from highly
excited MBL eigenstates in the topologically trivial region. For the topological Ising paramagnet in $d=2$ this last step still needs to be taken. In both cases we have argued the absence of an eigenstate phase transition separating the regions when the
preserving symmetry is allowed to be broken.

Evidently it would be interesting to extend this investigation to the larger zoo of
SPT phases identified in recent work, including in $d=3$ where SPT order can
presumably survive to non-zero temperatures when the disordering defects have the
topology of vortex lines. One immediate restriction suggested by our
analysis is that we found it necessary to protect the Haldane phase via a
discrete symmetry to invoke MBL. If that restriction is fundamental, it may be
that SPT order is strengthened by MBL only if the protecting symmetry is discrete.

\section{Acknowledgements}
We thank David Huse for very useful discussions on the question of many body localization of SPT phases with continuous symmetries, and for his valuable comments on a draft of this article. We would also like to acknowledge him and Vadim Oganesyan more broadly for multiple enlightening discussions on MBL over several years. We also thank M. Zaletel for drawing our attention to the effects of time reversal symmetry on the TPM phase. SLS would also like to acknowledge these gentlemen along with Rahul Nandkishore and Arijeet Pal for previous collaborative work that served as the inspiration for this project and Ehud Altman for a related discussion.
We acknowledge support by NSF Grant Numbers DMR 10-06608, PHY-1005429 (AC, VK and SLS), the John Templeton Foundation (SLS), the Lawrence Golub Fellowship (CRL), the NSF through a grant for ITAMP at Harvard University (CRL) and the Perimeter Institute for Theoretical Physics (AC and CRL). Research at Perimeter Institute is supported by the Government of Canada through Industry Canada and by the Province of Ontario through the Ministry of Research and Innovation.

\begin{appendix}

\section{Entanglement spectrum of dihedral symmetric MPS without translational symmetry}
\label{Sec:AppES}

In Ref.~\onlinecite{Pollmann:2010ih}, Pollman and co-authors demonstrated the entanglement spectrum of a spatial cut diagnoses the two dihedral symmetric translationally invariant phases of integer spins in one dimension. In the topological/Haldane phase, they showed that the entanglement spectrum is exactly double degenerate in the thermodynamic limit, while in the trivial phase, it is not. The two phases persist in the absence of translational invariance. In this appendix, we show that the classification of the entanglement spectrum of the MPS also holds without translational symmetry.

Our approach and notation closely follows that in Ref.~\onlinecite{Pollmann:2010ih}. Consider an open chain of a spin system with integer spin $S$ in the thermodynamic limit. Let the wavefunction of the system have a MPS representation (as is the case for the ground state of the clean system or the highly excited MBL states in the dirty system). The canonical form of such an MPS in the standard pictorial notation is:
\begin{align}
	\label{Eq:MPSrep}
	\includegraphics{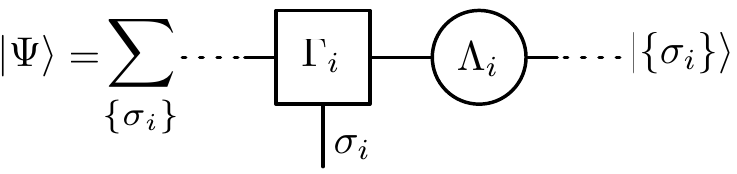}
\end{align}
where $i$ is the site label, $\sigma_i$ is the physical spin index taking values $-S, -S+1, \ldots S$, $\Gamma_i$ is a matrix of dimension $\chi$ and $\Lambda_i$ is a real, diagonal matrix, also of dimension $\chi$, with non-negative values. $\chi$ is interpreted as the dimension of the virtual spins that make up the spin $S$. \footnote{The proof may easily be extended to site-dependent $\chi$ ($\tilde{\chi}_i$). Then, the $\chi$ defined in the text is $\chi = \textrm{max} \tilde{\chi}_i$.} For a more detailed introduction to MPS, see \cite{Schollwock:2011qf, Verstraete:2008ko}. An important property of the canonical representation is that the transfer matrix at site $i$, defined as the tensor in the dashed box below, has a unique left (and right) eigenvector of eigenvalue one:
\begin{align}
	\label{Eq:TMatEig}
	\includegraphics{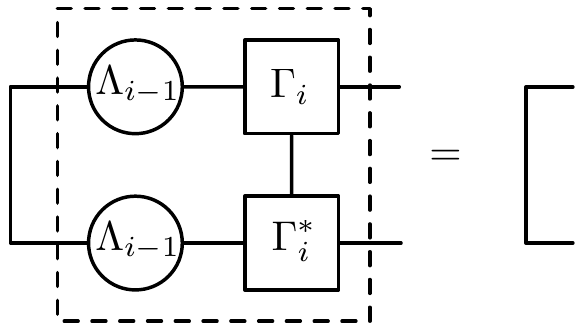}
\end{align}
The diagonal elements of $\Lambda_i$ are the Schmidt numbers for a spatial cut between bonds $i$ and $i+1$; the entanglement energies are the negative logarithms of these diagonal elements. Properties of the entanglement spectrum therefore follow from the structure of $\Lambda_i, \Gamma_i$.

To prove that there is a two-fold classification of the entanglement spectrum, we proceed as follows:
\begin{enumerate}
	\item Identify the action of the dihedral symmetry on the physical spins as a site-dependent gauge transformation of the virtual spins
	\item Show that the gauge transformation is the identity up to a site-dependent phase
	\item Determine that the smallest irreducible representation for the gauge transformation is either of dimension one, or two
\end{enumerate}
When the smallest irreducible representation has dimension two, the Schmidt values are forced to come in degenerate pairs and the entanglement spectrum is doubly degenerate. When the dimension is one, there is no constraint on the entanglement spectrum. This then is the required classification.

We now go through the steps in turn. Consider the action of the dihedral group on the state $\ket{\Psi}$. The matrix, $\Gamma_i$ in the MPS representation in Eq.~\eqref{Eq:MPSrep} becomes:
\begin{align}
	\tilde{\Gamma}_i^{\sigma} = (R_i^\alpha)^{\sigma\sigma'} \Gamma_i^{\sigma'}, \, \alpha=x,y,z
\end{align}
By definition, under the action of $\prod_i R_i^x$, $\prod_i R_i^y$ and $\prod_i R_i^z$, the given state goes back to itself, up to boundary effects that are not relevant in the thermodynamic limit. Thus, $\tilde{\Gamma}$ should be related to $\Gamma$ by a gauge transformation:
\begin{align}
\tilde{\Gamma}_i^{\sigma} = e^{i\theta_{i}^\alpha} (U^\alpha_{i-1})^\dagger \Gamma_i^{\sigma} U^\alpha_{i},
\end{align}
where $U_i^\alpha$ is a unitary matrix commuting with $\Lambda_i$ and $\theta_i^\alpha$ is real. Physically, the $U$ matrices implement the action of the symmetry on the virtual spins. They form a $\chi$-dimensional projective representation of the symmetry group of the wave function $\ket{\psi}$. Note that the MPS with matrices ($\tilde{\Gamma}_i^{\sigma}, \Lambda_i$) is also in the canonical representation. As the dihedral operators square to identity, another action of the dihedral group provides a relation for $\Gamma_i$:
 \begin{align}
	\label{Eq:GammaRel}
\Gamma_i^{\sigma} = e^{i2\theta_{i}^\alpha}\, (U^\alpha_{i-1})^\dagger (U^\alpha_{i-1})^\dagger \,\Gamma_i^{\sigma} \, U^\alpha_{i} U^\alpha_{i},
\end{align}

Substituting Eq.~\eqref{Eq:GammaRel} in Eq.~\eqref{Eq:TMatEig}, it is easily seen that:
\begin{align}
	\label{Eq:TmatEigVec}
	\includegraphics[width=8cm]{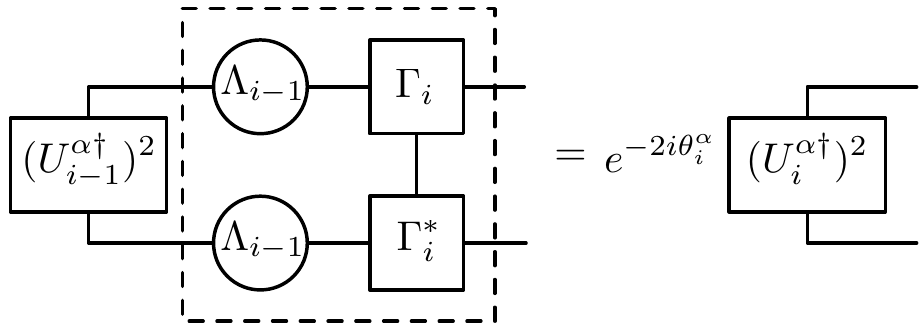}
\end{align}
The input left vector to the transfer matrix and the output vector are different. However, the norm of both vectors is $\chi$, equal to the norm of the unimodular eigenvector. As the transfer matrix has a unique unimodular eigenvector, both vectors have to be proportional to the identity eigenvector in Eq.~\eqref{Eq:TMatEig} up to a phase. Thus,
\begin{align}
	(U_{i-1}^{\alpha\dagger})^2 = e^{i\phi_{i-1}^\alpha} \mathbf{1}
\end{align}
This gets us to the second step in the list above. Further, as the eigenvalue is one, we obtain a relationship between $\theta^\alpha_i$, $\phi^\alpha_i$ and $\phi^\alpha_{i-1}$.

Finally, after a few steps of algebra, we find that:
\begin{align}
	(U_{i}^x)^\dagger (U_{i}^z)^\dagger &= \kappa (U_{i}^z)^\dagger(U_{i}^x)^\dagger \\
	\kappa &= \pm 1
\end{align}
That is, on every site $i$, $U_i^x$ and $U_i^z$ either commute or anti-commute. If $U_i^x$ and $U_i^z$ commute (anti-commute), the smallest irreducible representation has dimension one (two). Up to accidental degeneracies, $U_i^x, U_i^z$ can then be expressed as direct sums of matrices with dimension one (two). Recall however that $U_i^\alpha$ and the diagonal matrix with the Schmidt numbers, $\Lambda_i$, commute. Thus, in the former case, there is no constraint on the entanglement spectrum, while in the latter, the entire entanglement spectrum (ES) has to be doubly degenerate.

In the ground states of the clean/disordered A/BKLT chains, $\kappa=-1$ and the (ES) is two-fold degenerate. Consider now the fuzzy Fock states defined below Eq.~\eqref{Eq:AndersonTrip} using the localized single triplon wavefunctions, $\ket{\Psi} = t_{\psi_1}^{\alpha_1}t_{\psi_2}^{\alpha_2}\cdots t_{\psi_N}^{\alpha_N}\ket{A}$. In the extremely dilute limit, pick a bond $m$ where the weight of all the single triplon states occupied in $\ket{\Psi}$ is small. The local action of the dihedral group on this bond is the same as in the ground state and $\kappa=-1$ on this bond. As $\kappa$ is site-independent, $U_i^x$ and $U_i^z$ anti-commute for all $i$ and the entanglement spectrum will be doubly degenerate for any spatial cut. Thus, these approximate MBL states have the topological order of the Haldane phase.

\end{appendix}

\bibliography{papersCRL}

\end{document}